\documentclass[twocolumn,showpacs,preprintnumbers]{revtex4}%
\usepackage{amssymb}
\usepackage{amsmath}
\usepackage{epsfig}
\usepackage{amsfonts}
\usepackage{graphicx}%
\begin{document}
\title{Dynamical generation and dynamical reconstruction}
\author{Francesco Giacosa}
\affiliation{Institut f\"{u}r Theoretische Physik, Johann Wolfgang Goethe-Universit\"{a}t,
Max von Laue-Str.~1, D-60438 Frankfurt, Germany }

\begin{abstract}
A definition of `dynamical generation', a hotly debated topic at present, is
proposed and its implications are discussed. This definition, in turn, leads
to a method allowing to distinguish in principle tetraquark and molecular
states. The different concept of `dynamical reconstruction' is also introduced
and applies to the generation of preexisting mesons (quark-antiquark,
glueballs, ...) via unitarization methods applied to low-energy effective
Lagrangians. Large $N_{c}$ arguments play an important role in all these
investigations. A simple toy model with two scalar fields is introduced to
elucidate these concepts. The large $N_{c}$ behavior of the parameters is
chosen in order that the two scalar fields behave as quark-antiquark mesons.
When the heavier field is integrated out, one is left with an effective
Lagrangian with the lighter field only. A unitarization method applied to the
latter allows to `reconstruct' the heavier `quarkonium-like' field, which was
previously integrated out. It is shown that a Bethe-Salpeter (BS) analysis is
capable to reproduce the preformed quark-antiquark state, and that the
corresponding large-$N_{c}$ behavior can be brought in agreement with the
expected large $N_{c}$ limit; this is a subtle and interesting issue on its
own. However, when only the lowest term of the effective Lagrangian is
retained, the large $N_{c}$ limit of the reconstructed state is not
reproduced: instead of the correct large $N_{c}$ quarkonium limit, it fades
out as a molecular state would do. Implications of these results are
presented: it is proposed that axial-vector, tensor and (some) scalar mesons
just above 1 GeV obtained via the BS approach from the corresponding
low-energy, effective Lagrangian in which only the lowest term is kept, are
quarkonia states, in agreement with the constituent quark model, although they
might fade away as molecular states in the large $N_{c}$ limit.

\end{abstract}

\maketitle

\section{Introduction}

A central topic of past and modern hadron physics is the determination of the
wave function of resonances in terms of quark and gluon degrees of freedom,
both in the baryon and meson sectors and both for light and heavy quarks (for
reviews see refs. \cite{amslerrev,klempt}). In the mesonic sector, beyond
conventional quark-antiquark ($\overline{q}q$) mesons, one has glueball
states, multiquark states such as tetraquarks and `dynamically generated
resonances', most notably molecular states. Indeed, different authors used the
term `dynamical generation' in rather different contexts. In the works of
Refs. \cite{iam,daniel,av,te} a dynamical generated state is regarded as a
resonance obtained via unitarization methods from a low-energy Lagrangian; in
Refs. \cite{pelaeznc,leupoldrho,avnc} it is considered as a state which does
not follow the quark-antiquark pattern in large-$N_{c}$ expansion. In Refs.
\cite{pennington,morgan,tornqvist,vb} the concept of `additional, companion
poles' as dynamically generated states is introduced, while in Refs.
\cite{mesonicmol1,lemmer,speth} a dynamically generated resonance is regarded
as loosely bound molecular state. These various definitions are not mutually
exclusive and describe different points of view of the problem.

In this article (Sec. II and III) a definition for a dynamically generated
state is proposed and its implications, also in connection with the
aforementioned works, are presented. This definition, in turn, leads to a
method allowing to distinguish in principle tetraquark and molecular states,
although they are both four-quark states. The concept of `dynamical
reconstruction' is then introduced and discussed: it applies to resonances
which are obtained from low-energy effective Lagrangians via unitarization
methods, but still correspond to `fundamental' (not dynamically generated)
$\overline{q}q,$ glueball or multiquark states. In this context, the study of
the large $N_{c}$ behavior of these resonance constitutes a useful tool to
discuss their nature. In the end of Sec. III some general thoughts about the
form of an effective theory of hadrons valid up to 2 GeV are also presented.

In Sec. IV the attention is focused on a simple toy model, in which only two
scalar fields are considered. The parameters of the toy model are chosen in a
way that both fields behave as quarkonium-like state in the large $N_{c}$
limit. The heavier state is first integrated out in order to obtain an
effective low-energy Lagrangian in the toy world and then is reobtained via a
Bethe-Salpeter (BS) study applied to the low-energy Lagrangian. It is shown
that this is (at least in some cases) possible and that the corresponding
large $N_{c}$ behavior can be brought in agreement with the expected large
$N_{c}$ limit; this is a subtle and interesting issue on its own. However,
when only the lowest term of the effective, low-energy Lagrangian is retained,
the large $N_{c}$ limit of the reconstructed state is not reproduced: instead
of the correct large $N_{c}$ quarkonium limit (which must hold by
construction), it fades out as a molecular state would do. Implications of
these results are presented: it is proposed that axial-vector, tensor and
(some) scalar mesons just above 1 GeV obtained via the BS approach from the
corresponding low-energy, effective Lagrangian in which only the lowest term
is kept, are (reconstructed) quark-antiquark fields, in agreement with the
constituent quark model, although they might fade away as molecular states in
the large $N_{c}$ limit. Finally, in Sec. V the conclusions are presented.

\section{Dynamical generation}

Consider a physical system which is correctly and completely described in the
energy range $0\leq E\lesssim E_{\text{max}}$ by a quantum field theory in
which N fields $\phi_{1},$ $\phi_{2},...\phi_{N},$ their masses and
interactions are encoded in a Lagrangian $\mathcal{L}=\mathcal{L}(\phi
_{i},E_{\text{max}})$. Each Lagrangian has such an $E_{\text{max}}$ beyond
which it cannot be trusted. In particular, we shall refer to $E_{\text{max}}$
in the following sense: All the masses $M_{i}$ of the states $\phi_{i}$ and
the energy transfer in a two-body scattering $\phi_{i}$ $+\phi_{j}%
\rightarrow\phi_{i}+\phi_{j}$ should be smaller than $E_{\text{max}}$
$(M_{i}<E_{\text{max}}$ and, in the s-channel, $\sqrt{s}\lesssim E_{\text{max
}}$).

Moreover, we also assume that: (i) The theory is not confining, thus, to each
field $\phi_{i}$ there is a corresponding, measurable resonance (and at least
one with zero width). (ii) If not renormalizable, an appropriate
regularization shall be specified.

A resonance $R,$ emerging in the system described by $\mathcal{L}$, is said to
be dynamically generated if it does \emph{not} correspond to any of the
original fields $\phi_{1},$ $\phi_{2},...\phi_{N}$ $\subset$ $\mathcal{L}$ in
the Lagrangian and if its mass $M_{R}$ lies below $E_{\text{max}}$
($M_{R}\lesssim E_{\text{max}}$).

The last requirement $M_{R}\lesssim E_{\text{max}}$ is natural because the
state $R$ can be regarded as an additional, dynamically generated resonance
only if it belongs to the energy range in which the theory is valid. This
simple consideration plays an important role in the following discussion.
Clearly, the dynamical generated state $R$ emerges via interactions of the
original resonances $\phi_{i}.$ When switching them off, $R$ must disappear.
For this reason a dynamically generated mesonic resonance in QCD fades out in
the large-$N_{c}$ limit, which corresponds to a decreasing interaction
strength of mesons, see later for more details.

Some examples and comments are in order:

(a) $\mathcal{L=L}_{\text{QED}}$, in which the electron and the photon fields
are the basic fields. This theory is valid up to a very large $E_{\text{max}}$
(GUT scale). Positronium states are molecular, electron-positron bound states.
They appear as poles close to the real axis just below the threshold $2m_{e}$,
but slightly shifted due to their nonzero decay widths into photons. Clearly,
positronium states are `dynamically generated' according to the given
definition and should not be included in the original QED Lagrangian,
otherwise they would be double-counted. Note, the number of positronium states
is infinite.

(b) In Lagrangians describing nucleon-nucleon interaction via meson exchange
($\omega,$ $\rho,$ $\pi$ and $\sigma$) a bound state close to threshold,
called the deuteron, emerges via Yukawa-interactions, see for instance ref.
\cite{deut} and refs. therein. The deuteron is a dynamically generated
molecular state. In this case, when lowering the interaction strength below a
critical value (by reducing the coupling and/or increasing the mass of the
exchanged particle(s)), the bound state disappears. In fact, the number of
molecular states which can be obtained via a Yukawa potential is finite
\cite{yuk}, and eventually zero if the attraction is too weak. In such models
the deuteron should not be included in the original Lagrangian in order to
avoid double-counting \cite{fn1}.

(c) $\mathcal{L=L}_{\text{F}}$ the Fermi theory of the weak interaction, in
which the neutrino and electron fields interact via a local, quartic
interaction. This theory is valid up to $E_{\text{max}}<<M_{W},$ where $W$ is
the boson mediator of the weak force. As already mentioned in ref.
\cite{klempt}, the linear rise of the $\overline{\nu}_{e}e^{-}$ cross section
-as calculated from $\mathcal{L}_{\text{F}}$- shows a loss of unitarity at
high energy. Unitarization applied to $\mathcal{L}_{\text{F}}$ implies that a
resonance well above $E_{\text{max}}$ exists, and this resonance is exactly
the $W$ meson. However, being $M_{W}>E_{\text{max}}$ one cannot state in the
framework of the Fermi theory if the $W$ meson is dynamically generated or
not. A straightforward way to answer this question is the knowledge of the
corresponding theory valid up to an energy $E_{\text{max}}>M_{W}.$ Of course,
this theory is known: it is the electroweak theory described by the Lagrangian
$\mathcal{L}_{\text{EW}}$, which is part of the Standard Model \cite{djouadi}
and is valid up to a very high energy (GUT scale). In the framework of
$\mathcal{L}_{\text{EW}}$ the neutrino, the electron and the $W$ meson are all
elementary fields. One can then conclude that the $W$ meson is not a
dynamically generated state.

Indeed, $\mathcal{L}_{\text{F}}$ can be seen as the result of integrating out
the $W$ field from the electroweak Lagrangian $\mathcal{L}_{\text{EW}}$.
Unitarization arguments applied to the Fermi Lagrangian $\mathcal{L}%
_{\text{F}}$ allow in a sense to `dynamically reconstruct' the $W,$ which is
already present as a fundamental field in $\mathcal{L}_{\text{EW}}$.

(d) It is important to discuss in more depth and to formalize the issue raised
in the previous example. To this end let us consider the Lagrangian
$\mathcal{L}=\mathcal{L}(\phi_{i},E_{\text{max}})$ as the low-energy limit of
a Lagrangian $\mathcal{L}^{\prime}=\mathcal{L}^{\prime}(\phi_{i},\varphi
_{k},E_{\text{max}}^{\prime})$ valid up to an energy $E_{\text{max}}^{\prime
}>E_{\text{max}}.$ Beyond the fields $\phi_{i},$ $\mathcal{L}^{\prime}$
depends also on the fields $\varphi_{k}$, which are heavier than
$E_{\text{max}}$. Formally, when integrating out the fields $\varphi_{k}$ from
$\mathcal{L}^{\prime}$ one obtains $\mathcal{L}$.

In general, a unitarization scheme uses the information encoded in a
low-energy effective Lagrangian and the principle of unitarity in quantum
field theories, in order to deduce the existence and some properties of
resonances beyond the limit of validity of the theory itself. When applying a
unitarization scheme to the Lagrangian $\mathcal{L}$, an energy window between
$E_{\text{max }}$and a new energy scale $E_{U}>E_{\text{max }}$ - which
depends on the detail of the unitarization- becomes (partially) accessible.
For our purposes, we assume that $E_{U}\lesssim E_{\text{max}}^{\prime}$.

Let $R$ be a resonance with mass $E_{\text{max }}<M_{R}<E_{U}$ obtained from
$\mathcal{L}$ via a unitarization approach. Is this resonance $R$ dynamically
generated or not? A straightforward way to answer this question would be the
knowledge of $\mathcal{L}^{\prime}$. If $R$ corresponds to one of the fields
$\varphi_{k}$ is not dynamically generated and vice-versa. However, if
$\mathcal{L}^{\prime}$ is not known it is not possible to answer this question
at the level of the unitarized version of $\mathcal{L}$ only.

In conclusion, although the unitarization approach opens a window between
$E_{\text{max }}$and $E_{U}$ and the existence of resonances in this range can
be inferred from the unitarized Lagrangian $\mathcal{L}$ only, still the
knowledge of the latter is not complete \cite{fnlast}. If $\mathcal{L}%
^{\prime}$ is unknown, some other kind of additional information is required
to deduce the nature of $R$. In the framework of low-energy QCD this
additional information can be provided by large-$N_{c}$ arguments, see Sec. III.C.

(e) Let us consider a scalar field $\varphi=\varphi(t,x)$ in a 1+1 dimensional
world $(t,x)$ subject to the potential $V(\varphi)=\frac{\lambda}{4!}%
(\varphi^{2}-F^{2})^{2}.$ We assume that this theory is valid up to high
energies. When expanding around one of the two minima $\varphi=\pm F,$ the
mass of $\varphi$ is found to be $m=\lambda F^{2}/3.$ In addition, this theory
admits also a soliton with mass $M=\frac{2m^{3}}{\lambda},$ which is large if
$\lambda$ is small \cite{thooftlect}. In this example the solitonic state with
mass $M$ can be regarded as a `dynamically generated state'.

(f) Mixing can take place among two `fundamental fields' $\phi_{i}$ and
$\phi_{k}$: Two physical resonances arise as an admixture of these two fields.
One is predominantly $\phi_{i}$ and the other predominantly $\phi_{k}.$ Also,
mixing can take place among a dynamically generated resonance $R$ and one (or
more) of the $\phi_{i}.$ It decreases when the interaction is lowered
(large-$N_{c}$ limit in the mesonic world): one state reduces to the original,
preexisting resonance and the other disappears. In conclusion, mixing surely
represents a source of technical complication which renders the identification
of states (extremely) more difficult, but it does \emph{not }change the number
of states and the meaning of the previous discussion.

\section{Application to mesons}

\subsection{Effective hadron theory up to 2 GeV}

Let us turn to the hadronic world below 2 GeV. The basic ingredients of each
low-energy hadronic Lagrangian are quark-antiquark mesons and three-quark
baryons. In the framework of our formalism, we shall consider each
quark-antiquark (3-quark) state as a fundamental state, which is described by
a corresponding field in the hadronic Lagrangian (as long as its mass is below
an upper energy $E_{\text{max}}$).

Let us formalize this point in the mesonic sector as it follows. Consider the
correct, effective theory describing mesons up to $E_{\text{max}}\simeq2$ GeV
given by
\begin{equation}
\mathcal{L}_{eff}^{\text{had}}(E_{\text{max}},N_{c})%
\begin{array}
[c]{c}%
,
\end{array}
\label{1}%
\end{equation}
where $N_{c}$ is the number of colors. Its precise form is, unfortunately,
unknown. In fact, because confinement has not yet been analytically solved, it
is not possible to derive $\mathcal{L}_{eff}^{\text{had}}(E_{\text{max}}%
,N_{c})$ from the QCD Lagrangian. In the limit $N_{c}\rightarrow\infty$ the
effective Lagrangian $\mathcal{L}_{eff}^{\text{had}}(E_{\text{max}},N_{c})$ is
expected to be more simple. Although even in this limit a mathematical
derivation is not possible, it is known that it must primarily consists of
non-interacting quark-antiquark states. In fact, their masses scale as
$N_{c}^{0}$ and their decay widths as $N_{c}^{-1}$ respectively
\cite{thooft,witten,coleman,lebed}. Considering that the $\overline{q}q$ mass
scales as $N_{c}^{0},$ these states shall be clearly present also when going
back from the large $N_{c}$ limit to the physical world $N_{c}=3.$ The
next-expected states which are present in the large $N_{c}$ limit are
glueballs, i.e. bound states of pure gluonic nature. Their masses also scale
as $N_{c}^{0}$ and the decay widths as $N_{c}^{-2}.$ They thus are also
expected to be present in the real world for $N_{c}=3.$In particular, the
lightest glueball is a scalar field which is strongly related to the trace
anomaly, i.e. the breaking of the classical dilatation invariance of the QCD
Lagrangian (see also Sec. III.D). In addition to quark-antiquark and glueball
states, also hybrid states survive in the large-$N_{c}$ limit \cite{lebed}.
They constitute an interesting subject of meson spectroscopy (see ref.
\cite{hybrid} and refs. therein), but will not be considered in the following
discussion. All these states are therefore `preexisting' and not-dynamically
generated states of the mesonic Lagrangian under consideration.

An intermediate comment is devoted to baryon states: they have an linearly
increasing mass with $N_{c}$ ($M\sim N_{c}$) which exceeds $E_{\text{max}}$
for a large enough $N_{c}$ and are therefore not present in the large-$N_{c}$
limit of $\mathcal{L}_{eff}^{\text{had}}(E_{\text{max}},N_{c})$. Thus,
although they appear in the $N_{c}=3$ world, they are not present in the
effective Lagrangian because of the way in which the limit is constructed. If,
instead, we construct the large $N_{c}$ limit as $\mathcal{L}_{eff}%
^{\text{had}}(\frac{N_{c}}{3}E_{\text{max}},N_{c}\rightarrow\infty)$ (in such
a way that at $N_{c}=3$ it coincides with Eq. (\ref{1})), baryons are well
defined states as proved originally in ref. \cite{witten}. For simplicity
baryons will not be discussed in this paper, but together with hybrid states,
should be included in a more complete treatment.

The reason why the value $E_{\text{max}}\simeq2$ GeV is chosen is that all the
resonances under study in this work are lighter than 2 GeV. Thus, they either
correspond to a field in $\mathcal{L}_{eff}^{\text{had}}(E_{\text{max}}%
,N_{c}=3)$ or arise as additional resonances (i.e. dynamically generated) via
interaction of preexisting states of $\mathcal{L}_{eff}^{\text{had}%
}(E_{\text{max}},N_{c}=3)$. The full knowledge of the Lagrangian
$\mathcal{L}_{eff}^{\text{had}}(E_{\text{max}}\simeq2$ GeV$,N_{c}=3)$ would
allow to answer if a resonance lighter than 2 GeV is dynamically generated or
not by a simple look at it. Clearly, if one would be interested in a resonance
whose mass is heavier than 2 GeV, then $E_{\text{max}}$ should be increased.
Moreover, one expects to find below 2 GeV all the relevant ground-state mesons
in the channel $J^{PC}=0^{-+},$ $0^{++},$ $1^{-+}$, $1^{++},$ $2^{++}$. Thus,
$\mathcal{L}_{eff}^{\text{had}}(2$ GeV$,3)$ may be described by an effective
Lagrangian which exhibits linear realization of chiral symmetry and its
spontaneous breakdown, see section III.D for a closer discussion.

Most of the mesonic resonances listed in the Particle Data Group \cite{pdg}
can be immediately associated to a corresponding, underlying quark-antiquark
state. Yet, the question if some resonances of ref. \cite{pdg} are not
$\overline{q}q$ is interesting and at the basis of many studies. In the
mesonic sector, two alternative possibilities are well-known:

$\bullet$ Molecular states: they are bound state of two distinct
quark-antiquark mesons. They correspond to the example of the positronium
(example (a) Sec. II). Just as the positronium states are not included in the
QED Lagrangian, hadronic molecular states should \emph{not} be included
directly in $\mathcal{L}_{eff}^{\text{had}}(E_{\text{max}},N_{c}).$ They arise
upon meson-meson interactions in the $N_{c}=3$ physical world, see the general
discussion above. However, they inevitably fade out in the large-$N_{c}$ limit
because the interaction of a n-leg meson vertex decreases as $N_{c}%
^{-(n-2)/2}.$ This is therefore a clear example of dynamically generated
states within a mesonic system (see also the next subsection for a closer
description of physical candidates below 1 GeV).

$\bullet$ Tetraquark states: they consist of two distinct, colored `bumps', in
contrast to a molecular state, which is made of two colorless, quark-antiquark
bumps \cite{fn2}. Loops of $\overline{q}q$ mesons, corresponding to the
interaction of two colorless states, cannot generate the color distribution of
a tetraquark. If present at $N_{c}=3$, they shall be included directly in the
effective Lagrangian $\mathcal{L}_{eff}^{\text{had}}(E_{\text{max}},N_{c}=3)$
and, in view of the given definition, should not be regarded as dynamically
generated states.

A second, slightly different way to see it is the following: let us imagine to
construct the Lagrangian $\mathcal{L}_{eff}^{\text{had}}(E_{\text{max}}%
,N_{c}).$ We first put in quark-antiquark and glueball states, that is those
configurations which surely survive in the large-$N_{c}$ limit and corresponds
to non-dynamically generates states: $\mathcal{L}_{eff}^{\text{had}%
}(E_{\text{max}},N_{c})=\mathcal{L}_{eff}^{\overline{q}q\text{+glueballs}%
}(E_{\text{max}},N_{c}).$ Then the question is: `does this Lagrangian describe
the physical world for $N_{c}=3$'? (Note, loops shall be taken into account
and dynamically generated states can eventually emerge out of this
Lagrangian.) If the answer is positive, no multiquark states are needed. If
the answer is negative, the basic Lagrangian shall be extended to include from
the very beginning multiquark states, most notably tetraquark states:
$\mathcal{L}_{eff}^{\text{had}}(E_{\text{max}},3)=\mathcal{L}_{eff}%
^{\overline{q}q\text{+glueballs}}(E_{\text{max}},3)+\mathcal{L}_{eff}%
^{\text{multiquark}}(E_{\text{max}},3).$

A third approach to the problem goes via large-$N_{c}$ arguments. In refs.
\cite{witten,coleman} it has been shown that a tetraquark state also vanishes
in the large-$N_{c}$ limit. However, for $N_{c}=3$ the most prominent and
potentially relevant for spectroscopy is the `good' diquark, which is
antisymmetric in color space: $d_{a}=\varepsilon_{abc}q^{b}q^{c},$ (with
$a,b,c=1,2,3)$ \cite{exotica}. The tetraquark is the composition of a good
diquark and a good antidiquark: $d_{a}^{\dagger}d_{a}$. The extension to
$N_{c}$ of a good diquark is the antisymmetric configuration $d_{a_{1}%
}=\varepsilon_{a_{1}a_{2}a_{3}...a_{N_{c}}}q^{a_{2}}q^{a_{3}}...q^{a_{N_{c}}}$
with $a_{1},...a_{N_{c}}=1,...N_{c},$ which constitutes of $(N_{c}-1)$ quarks.
Thus, the generalization of the tetraquark to the $N_{c}$ world is not a
diquark-antidiquark object, but the state $\chi=\sum_{a_{1}=1}^{N_{c}}%
d_{a_{1}}^{\dagger}d_{a_{1}}$ which is made of $(N_{c}-1)$ quarks and
$(N_{c}-1)$ antiquarks, see also the discussion in ref. \cite{liu}. It is the
dibaryonium already described in ref. \cite{witten} which has a well defined
large-$N_{c}$ limit: its mass scale as $M_{\chi}\varpropto2(N_{c}-1)$ and
decays into a baryon and an antibaryon. The state $\chi,$ while not present in
$\mathcal{L}_{eff}^{\text{had}}(E_{\text{max}},N_{c}\rightarrow\infty)$
because its mass overshoots $E_{\text{max}}$, appears in $\mathcal{L}%
_{eff}^{\text{had}}(N_{c}E_{\text{max}},N_{c}\rightarrow\infty)$ in which also
the baryons survive: this is contrary to a dynamically generated state, which
disappears also in this case.

As a result of our discussion, tetraquark states and molecular states,
although both formally four-quark states, are crucially different: The former
are `elementary' and should be directly included in the effective Lagrangian
$\mathcal{L}_{eff}^{\text{had}}(E_{\text{max}},N_{c}=3)$, the latter can
emerge as dynamically generated resonances. We now turn to the particular case
of the light scalar mesons, where all these concepts play an important role.

\subsection{Light scalar mesons}

One of the fundamental questions of low energy QCD concerns the nature of the
lightest scalar states $\sigma\equiv f_{0}(600),$ $k\equiv k(800),$
$f_{0}\equiv f_{0}(980)$ and $a_{0}\equiv a_{0}(980).$ Shall these states be
included from the very beginning in $\mathcal{L}_{eff}^{\text{had}%
}(E_{\text{max}},N_{c})$? If yes, they correspond to quark-antiquark or
tetraquark nonets (one of them can be also related to a light scalar
glueball). If not, they shall be regarded as dynamically generated states. The
main point of the following subsection is to discuss previous works about
light scalar mesons in connection with the proposed definition of dynamical
generation. In fact, it is easy to classify previous works into two classes
(not dynamically generated and dynamically generated), thus allowing to order
different works of the last three decades in a clear way. We first review
works in which scalar states are not dynamically generated and then works in
which they are dynamically generated.

\emph{`The light scalar states are not dynamically generated and should be
directly included in }$\mathcal{L}_{eff}^{\text{had}}(E_{\text{max}},N_{c}%
)$\emph{':}

(i) In the quark-antiquark picture these light scalar states form the nonet of
chiral partners of pseudoscalar mesons. Their flavor wave functions reads
$\sigma\simeq\sqrt{\frac{1}{2}}(\overline{u}u+\overline{d}d),$ $f_{0}%
\simeq\overline{s}s,$ $a_{0}^{+}\equiv u\overline{d},$ $k^{+}\equiv
u\overline{s}.$ At the microscopic level this is the prediction of the NJL
model \cite{njl}, where a $\sigma$ mass of about $2m^{\ast}$ is obtained and
where $m^{\ast}\sim300$ MeV is the constituent quark mass. This is usually the
picture adopted in linear sigma models at zero \cite{scadron} and at nonzero
density and temperature \cite{lenaghan}. However, this assignment encounters a
series of problems: it can hardly explain the mass degeneracy of $a_{0}$ and
$f_{0},$ the strong coupling of $a_{0}$ to kaon-kaon, the large mass
difference with the other p-wave nonets of tensor and axial-vector mesons
\cite{godfrey}, it is at odd with large-$N_{c}$ studies (see later on) and
with recent lattice works \cite{mathur}.

(ii) Tetraquark picture, first proposed by Jaffe \cite{jaffeorig} and
revisited in Refs. \cite{maiani,bugg,tq}: $\sigma\simeq\frac{1}{2}%
[\overline{u},\overline{d}][u,d],$ $f_{0}\simeq\frac{1}{2\sqrt{2}}%
([\overline{u},\overline{s}][u,s]+[\overline{d},\overline{s}][d,s]),$
$a_{0}^{+}\equiv\frac{1}{2}[\overline{d},\overline{s}][u,d],$ $k^{+}%
\equiv\frac{1}{2}[\overline{d},\overline{s}][u,d]$ where $[.,.]$ stands for
antisymmetric configuration in flavor space (which, together with the already
mentioned antisymmetric configuration in color space, implies also a s-wave
and spinless structure of the diquarks and of the tetraquarks). Degeneracy of
$a_{0}$ and $f_{0}$ is a natural consequence. A good phenomenology of decays
can be obtained if also the next-to-leading order contribution in the
large-$N_{c}$ expansion is taken into account \cite{tq} and/or if instanton
induced terms are included \cite{thooftlast}. Linear sigma models with an
additional nonet of scalar states can be constructed
\cite{fariborz,napsuciale,tqmix}. The quark-antiquark states lie above 1 GeV
\cite{refs} and mix with the scalar glueball whose mass is placed at $\sim1.7$
GeV by lattice QCD calculations \cite{lattglue}. This reversed scenario
directly affects the physics of chiral restoration at non-zero temperature
\cite{achim}.

(iii) Different assignments, in which also the glueball state shows up below 1
GeV have been proposed, see refs. \cite{klempt,minkowski,vento} and refs. therein.

(iv) In all these assignments the very existence of the scalar mesons is due
to some preformed compact bare fields entering in $\mathcal{L}_{eff}%
^{\text{had}}(E_{\text{max}},N_{c}=3)$. By removing the corresponding bare
resonances from $\mathcal{L}_{eff}^{\text{had}}(E_{\text{max}},N_{c}=3),$ they
disappear. Dressing via meson-meson loops, such as $\pi\pi$ for $\sigma,$
$K\pi$ for $k$ and $KK$ for $a_{0}$ and $f_{0},$ surely takes place. In
particular, due to the intensity in these channels and the closeness to
thresholds, they can cause a strong distortion and affect the properties of
the scalar states \cite{hanhart}. However, the important point is that in all
these scenarios mesonic loops represent a further complication of light
scalars, but are not the reason of their existence.

(v) As discussed in Section III.A, non-dynamically generated scalar states
survive in the large-$N_{c}$ limit, although in a different way according to
quarkonium, glueball or tetraquark interpretations.

\emph{`The light scalars states are dynamically generated and} \emph{should
not be included in }$\mathcal{L}_{eff}^{\text{had}}(E_{\text{max}},N_{c})$':

(i) In Ref. \cite{speth} the $\sigma$ pole arises as a broad enhancement due
to the inclusion of $\rho$ mesons in the t-channel isoscalar $\pi\pi$
scattering. In this case the $\sigma$ is `dynamically generated' and arises
because of a Yukawa-like interaction due to $\rho$ meson exchange (pretty much
as the deuteron described above, but above threshold). When reducing the
$\rho\pi\pi$ coupling $g_{\rho\pi\pi}$ (which, in the large-$N_{c}$ limit
scales as $1/\sqrt{N_{c}}$), the $\sigma$ fades out. Alternatively, the limit
$M_{\rho}\rightarrow\infty$ also implies a disappearance of the $\sigma$-enhancement.

(ii) Similar conclusions for the $f_{0}(980)$ and $a_{0}(980)$ meson,
described as molecular $\overline{K}K$ bound states just below threshold, have
been obtained in refs. \cite{mesonicmol1}. In particular, in ref.
\cite{lemmer} the origin of these states is directly related to a
one-meson-exchange potential. Within all these approaches the $a_{0}(980)$ and
the $f_{0}(980)$ are `dynamically generated'.

(iii) In the model of ref. \cite{pennington} the $a_{0}$ state also arises as
an additional, dynamically generated state, but in a different way. Scalar and
pseudoscalar quark-antiquark mesons are the original states. A bare scalar
state with a mass of $1.6$ GeV is the original, quark-antiquark `seed'. When
loops of pseudoscalar mesons are switched on, the mass is slightly lowered and
the state is identified with $a_{0}(1450).$ In addition, a second state,
arising in this model as a further zero of the real part of the denominator of
the propagator, is identified with the $a_{0}(980)$ meson: it is dynamically
generated and disappears in the large-$N_{c}$ limit, where only the original
quark-antiquark seed survives. More in general, we refer to
\cite{tornqvist,morgan,vb} for the emergence of additional, companion poles
not originally present as preexisting states in the starting Lagrangian. In
particular, in ref. \cite{vb} the conventional scalar quark-antiquark states,
calculated within an harmonic oscillator confining potential, lie above 1 GeV.
When meson loops are switched on, a \emph{complete} second nonet of
dynamically generated states below 1 GeV emerge. Note, in all these studies
the validity of the employed theories lies well above 1 GeV, so that the
definition of dynamical generation given in Section II holds for the light scalars.

(iv) Note, in (i) and (ii) the emergence of states is due to t-channel forces.
This is not the case in (iii). However, a common point of them is that the
light scalar states disappear in the large-$N_{c}$ limit.

It is clear that the situation concerning light scalars is by far not
understood. We wish, however, to stress once more that there is a crucial
difference among the two outlined options in relation to the Lagrangian
$\mathcal{L}_{eff}^{\text{had}}(E_{\text{max}},N_{c}).$ Note also that in this
subsection we only discussed works for which it is possible to immediately
conclude if the scalar mesons are dynamically generated or not according to
the definition given in Sec. II. Unitarization methods were not discussed
here; in fact, when the latter are applied, care is needed. This is the
subject of the next subsection.

\subsection{Low-energy Lagrangians, unitarization and dynamical
reconstruction}

The Lagrangian\textbf{\ }$\mathcal{L}_{eff}^{\text{had}}(E_{\text{max}}%
,N_{c})$ with $E_{\text{max}}\simeq2$ GeV induces breakdown of chiral symmetry
$SU_{A}(N_{f}),$ where $N_{f}$ is the number of light flavors. There are
therefore $N_{f}^{2}-1$ Goldstone bosons: the pion triplet for $N_{f}=2,$ in
addition four kaonic states and the $\eta$ meson for $N_{f}=3.$

If we integrate out all the fields in $\mathcal{L}_{eff}^{\text{had}%
}(E_{\text{max}},N_{c})$ besides the three light pions, we obtain the
Lagrangian of chiral perturbation theory (see ref. \cite{scherer} and refs.
therein) for $N_{f}=2:$%
\begin{equation}
\mathcal{L}_{eff}^{\text{had}}(E_{\text{max}},N_{c})\rightarrow\mathcal{L}%
_{\chi PT}(E_{\chi PT},N_{c})%
\begin{array}
[c]{c}%
,
\end{array}
\end{equation}
where $E_{\chi PT}$ should be lower than the first resonance heavier than the
pions ($\sim400$ MeV). $\mathcal{L}_{\chi PT}(E_{\chi PT},N_{c})$ is recasted
in an expansion of the pion momentum $O(p^{2n})$ and for each $n$ there is a
certain number of low-energy coupling constants, which in principle could be
calculated from $\mathcal{L}_{eff}^{\text{had}}(E_{\text{max}},N_{c}),$ if it
were known. Being this not the case, they are directly determined by
experimental data. (Similar properties hold when the kaons and the $\eta$ are
retained in $\mathcal{L}_{\chi PT}(E_{\chi PT},N_{c})$).

For instance, the vector isotriplet $\rho$ meson is predicted by a large
variety of approaches (such as quark models) to be a preexisting $1^{--}$
quark-antiquark field. In this sense it is a fundamental field appearing in
$\mathcal{L}_{eff}^{\text{had}}(2$ GeV$,3),$ which is integrated out (together
with other fields) to obtain $\mathcal{L}_{\chi PT}(E_{\chi PT},3).$ However,
the $\rho$ meson spectral function cannot be obtained from chiral perturbation
theory unless a unitarization scheme is employed
\cite{daniel,iam,pelaeznc,leupoldrho}. As an example, via the IAM (Inverse
Amplitude Method) unitarization scheme applied to $\mathcal{L}_{\chi
PT}(E_{\chi PT},N_{c}=3)$ \cite{iam}, a window between the original energy
$E_{\chi PT}$ and $4\pi f_{\pi}\sim1$ GeV is opened: resonances with masses in
this window, such as the $\rho$ meson, can be described within unitarized
$\chi PT.$ As discussed in the point (d) of Sec. II, the very last question if
the $\rho$ meson is dynamically generated or not \emph{cannot} be answered at
the level of unitarized $\chi PT$ . One still does not know if $\rho$
corresponds to a basic, preexisting field entering in $\mathcal{L}%
_{eff}^{\text{had}}(2$ GeV, $N_{c})$ or not.

Some additional information is needed. In the interesting and important case
of large-$N_{c}$ studies of unitarized $\chi PT$, the required additional
knowledge is provided by the large-$N_{c}$ scaling of the low-energy
constants: It has been shown in ref. \cite{pelaeznc} that the $\rho$ mass
scales as $N_{c}^{0}$ and the width as $N_{c}^{-1}$ and thus the $\rho$ meson
should be considered as a fundamental (not dynamically generated)
quark-antiquark field, which shall be directly included in $\mathcal{L}%
_{eff}^{\text{had}}(E_{\text{max}},N_{c}=3)$. We also refer to the analytic
results of \cite{leupoldrho} where the large-$N_{c}$ limit is evident.

Let us turn to the lightest scalar-isoscalar resonance $\sigma\equiv
f_{0}(600)$ as obtained from (unitarized) $\chi PT$. In refs.
\cite{caprini,yndurain} precise determinations of the $\sigma$ pole are
obtained, but -as stated in ref. \cite{caprini}- it is difficult to understand
its properties in terms of quarks and gluons. In ref. \cite{pelaeznc} a study
of the $\sigma$ pole within the IAM-scheme in the large-$N_{c}$ limit has been
performed: a result which is at odd with a predominantly quarkonium, or a
glueball, interpretation of the $\sigma$ meson has been obtained. The mass is
not constant and the width does not decrease. However, even at this stage one
still cannot say if the $\sigma$ is `dynamically generated' or `reconstructed'
in relation to $\mathcal{L}_{eff}^{\text{had}}(E_{\text{max}},N_{c}=3)$
because it is hard to distinguish the molecular and the tetraquark assignments
in the large-$N_{c}$ limit (see discussion above).

Recently, the Bethe-Salpeter unitarization approach has been used to generate
various axial-vector \cite{av,avnc}, tensor and scalar mesons above 1 GeV
\cite{te}. The starting point are low-energy Lagrangians for the
vector-pseudoscalar (such as $\rho\pi$) and vector-vector (such as $\rho\rho)$
interactions. These Lagrangians are also in principle derivable by integrating
out heavier fields from the complete $\mathcal{L}_{eff}^{\text{had}%
}(E_{\text{max}},N_{c}=3)$. For instance, in the $\rho\pi$ axial-vector
channel the $a_{1}(1260)$ meson is obtained and in the $\rho\rho$ tensor and
scalar channels the states $f_{2}(1270)$ and $f_{0}(1370)$ are found. Are
these dynamically generated states of molecular type? The answer is: not
necessarily. In fact, the masses of the obtained states lie \emph{above }the
energy limit of the low-energy effective theories out of which they are
derived. Even for these states the possibility of `dynamical reconstruction'
-just as for the $\rho$ meson described above- is not excluded: In this
scenario, these resonances above 1 GeV are intrinsic, preexisting
quark-antiquark or glueball (multiquark states are improbable here) fields of
$\mathcal{L}_{eff}^{\text{had}}(E_{\text{max}},N_{c}=3).$ While first
integrated out to obtain the low-energy Lagrangians, unitarization methods
applied to the latter allow to reconstruct them. In the next section a toy
model is presented, in which this mechanism is explicitly shown: although a
state obtained via BS-equation looks like a molecule, it still can represent a
fundamental, preexisting quark-antiquark (or glueball) state.

As discussed in the summary of the PDG compilation \cite{pdgnote} and in refs.
\cite{amslerrev,klempt} (and refs. therein) the tensor resonances
$f_{2}(1275),$ $f_{2}(1525),$ $a_{2}(1320)$ and $K_{2}(1430)$ represent a
nonet of quark-antiquark states. The ideal mixing, the very well measured
strong and electromagnetic decay rates \cite{tensormio}, the masses and the
mass splitting are all in excellent agreement with the quark-antiquark
assignment. In this case they are fundamental (intrinsic) fields of
$\mathcal{L}_{eff}^{\text{had}}(E_{\text{max}},N_{c}=3),$ which can be
dynamically reconstructed (rather than generated) via unitarization scheme(s)
applied to a low-energy Lagrangians.

Although experimentally and theoretically more involved, the same can hold in
the axial-vector channel: the resonances $f_{1}(1285),$ $f_{1}(1510),$
$a_{1}(1260)$ and $K_{1}(1270)$ are in good agreement with the low-lying
$1^{++}$ quark-antiquark assignment. Even more complicated is the situation in
the scalar channel: the low-lying quark-antiquark states mix with the scalar
glueball \cite{refs}. Also in this case, however, the possibility of
`dynamical reconstruction' rather then `generation' is upheld.

If dynamical reconstruction takes place, there is no conflict between the
quark model assignment of ref. \cite{pdgnote} and the above mentioned recent
studies. Note, also, that dynamical reconstruction is in agreement with the
discussion of ref. \cite{gasser}.

\subsection{A simplification of the Lagrangian $\mathcal{L}_{eff}^{\text{had}%
}(E_{\text{max}},N_{c})$}

The Lagrangian $\mathcal{L}_{eff}^{\text{had}}(E_{\text{max}},N_{c})$ with
$E_{\text{max}}\sim2$ GeV has been a key ingredient allover the present
discussion but it has not been made explicit because unknown. An improved
knowledge of (at least parts of) $\mathcal{L}_{eff}^{\text{had}}%
(E_{\text{max}},N_{c})$ would surely represent a progress in understanding the
low-energy hadron system. As a last step we discuss which properties it might
have. Clearly, it must reflect the symmetries of QCD, most notably spontaneous
breakdown of chiral symmetry. The (pseudo-) scalar meson matrix $\Phi$, the
(axial-)vector, tensor mesons and the scalar glueball are its basic building
blocks. Moreover, if additional non-dynamically generated scalar states such
as tetraquarks exist, they shall be also included. We thus have a complicated,
general $\sigma$-model Lagrangian with many terms, in which operators of all
orders can enter, because renormalization is not a property that an effective
hadronic Lagrangian should necessarily have. The question is if it is possible
to obtain a (relatively) simple form out of this complicated picture, see also
\cite{fariborz,napsuciale,tqmix,farnew}.

A possibility to substantially simplify the situation is via dilation
invariance; let us consider the (pseudo-)scalar meson matrix $\Phi,$ which
transforms as $\Phi\rightarrow R\Phi L^{\dagger}$ ($R,L$ $\subset SU(3)$)
under chiral transformation and the dilaton field $G$, subject to the
potential \cite{salomone} $V_{G}=\propto G^{4}(\log G/\Lambda_{G}+1/4)$, where
$\Lambda_{G}$ is a dimensional parameter of the order of $\Lambda_{QCD}$ (the
glueball emerges upon shifting $G$ around the minimum of its potential
$G_{0}\sim\Lambda_{G}$). Consider $\mathcal{L}_{eff}^{\text{had}%
}(E_{\text{max}},N_{c})=T-V,$ where $T$ is the dynamical part and $V=V[G,$
$\Phi,...]$ is the potential describing masses and interactions of the fields
(dots refer to other degrees of freedom, such as (axial-)vector ones). We
assume that: (i) In the chiral limit the only term in $V$ which breaks
dilation invariance -and thus mimics the trace anomaly of QCD- is encoded in
$V_{G}$ (via the dimensional parameter $\Lambda_{G}$). (ii) The potential $V$
is finite for any finite value of the fields. As a consequence of (i) only
operators of order (exactly) four can be included. They have the form
$G^{2}Tr[\Phi^{\dagger}\Phi],$ $Tr[\Phi^{\dagger}\Phi\Phi^{\dagger}\Phi],$
$Tr[\Phi^{\dagger}\Phi]^{2},...$ As a consequence of (ii) a huge set of
operators are not admitted. In fact, an operator of the kind $G^{-2}%
Tr[\partial_{\mu}\Phi^{\dagger}\partial^{\mu}\Phi]^{2},$ is excluded because,
although of dimension 4, it blows up for $G\rightarrow0$. In this way we are
left with a sizably smaller number of terms, even smaller than what
renormalizability alone would impose \cite{fn5}. Work alone this direction,
including (axial-)vector degrees of freedom is ongoing \cite{frankfurt} and
can constitute an important source of informations for spectroscopy and for
future developments at nonzero temperature and densities, where in the
chirally restored phase a degeneration of chiral partners is manifest.

In conclusion, a way to implement these ideas and use the definition of
dynamical generation can be sketched as follows: after writing a general
chirally symmetric Lagrangian up to fourth order including the glueball and
the quark-antiquark fields as basic states, one should attempt without further
inclusion of any other state to describe physical processes up to $\sim2$ GeV,
as pion-pion scattering, decay widths, etc. In doing this one should of course
include loops. If, for instance, we start with a basic scalar-isoscalar
quark-antiquark field above 1 GeV, do we correctly reproduce the resonance
$f_{0}(600)$ when solving the Bethe-Salpeter channel in the $\pi\pi$ sector
below 1 GeV? If the answer is positive, the latter resonance is dynamically
generated and there is no need of any other additional state. If, albeit
including loops, the attraction among pions turns out to be too weak to
generate the resonance $f_{0}(600),$ we conclude that it is necessary to
enlarge our model by explicitly introducing a field which describes it. As
argued previously, this field can be identified as a tetraquark state. If this
ambitious program will lead to a successful result is matter of future research.

\section{A toy model for dynamical reconstruction}

\subsection{Definitions and general discussion}

In this section we start from a toy Lagrangian, in which two mesons $\varphi$
and $S$ interact. A large-$N_{c}$ dependence is introduced in such a way that
both fields behave as quarkonium states. Then, the field $S$, which is taken
to be heavier than $\varphi,$ is integrated out and a low-energy Lagrangian
with the field $\varphi$ only is obtained. A Bethe-Salpeter study is applied
to the latter Lagrangian: the question is if the original state $S$, which was
previously integrated out, can be reobtained in this way. The answer is
generally positive, however care is needed concerning the large-$N_{c}$ limit.
In a straight BS-approach the quarkonium-like large-$N_{c}$ limit of the state
$S$ cannot be reproduced. However, as it shall be shown, within a modified BS
approach the large-$N_{c}$ limit can be correctly obtained.

The toy Lagrangian \cite{lupo,lupoaltri} consisting of the two fields
$\varphi$ (with mass $m$) and $S$ (with bare mass $M_{0}>2m$) reads%
\begin{equation}
\mathcal{L}_{\text{toy}}(E_{\text{max}},N_{c})=-\frac{1}{2}\varphi
(\square+m^{2})\varphi-\frac{1}{2}S(\square+M_{0}^{2})S+gS\varphi^{2}%
\begin{array}
[c]{c}%
,
\end{array}
\label{ltoy}%
\end{equation}
which we assume to be valid up to $E_{\text{max}}>>M_{0}.$ The $N_{c}$
dependence of the effective Lagrangian is encoded in $g$ only: $g=g(N_{c}%
)=g_{0}\sqrt{3/N_{c}}.$ In this way both masses behave like $N_{c}^{0}$ and
the decay amplitude for $S\rightarrow2\varphi$ scales as $1/\sqrt{N_{c}},$
just as if $\varphi$ and \ $S$ were quarkonia states. $\mathcal{L}%
_{toy}(E_{\text{max}},N_{c}=3)$ is the analogous of $\mathcal{L}%
_{eff}^{\text{had}}(E_{\text{max}},N_{c}=3)$ in a simplified toy world. For
definiteness we refer to values in GeV: $m=0.3$, $M_{0}=1,$ $g_{0}$ will be
varied within $1.5$ and $5.$%

%TCIMACRO{\FRAME{ftbpFU}{3.7343in}{3.1626in}{0pt}{\Qcb{Eqs. (4), (6) and (9)
%are depicted: In eq. (4) the $S$ resonance is dressed via loops of $\varphi$
%mesons. In eq. (6) the $T$-matrix is represented by an exchange of a dressed
%$S$ meson. Finally, eq. (9) represents a BS equation applied to the quartic
%terms of the Lagrangian $\QTR{cal}{L}_{\text{le}}(E_{\text{le}},N_{c})$ is
%represented, in which the $T$ matrix appears both on the left and the right
%hand sides of the equation.}}{}{fignew.eps}%
%width 3.7343in;  height 3.1626in;  depth 0pt;  original-width 8.6844in;
%original-height 7.344in;  cropleft "0";  croptop "1";  cropright "1";
%cropbottom "0";  filename 'fignew.eps';file-properties "XNPEU";}} }%
%BeginExpansion
\begin{figure}
[ptb]
\begin{center}
\includegraphics[
height=3.1626in,
width=3.7343in
]%
{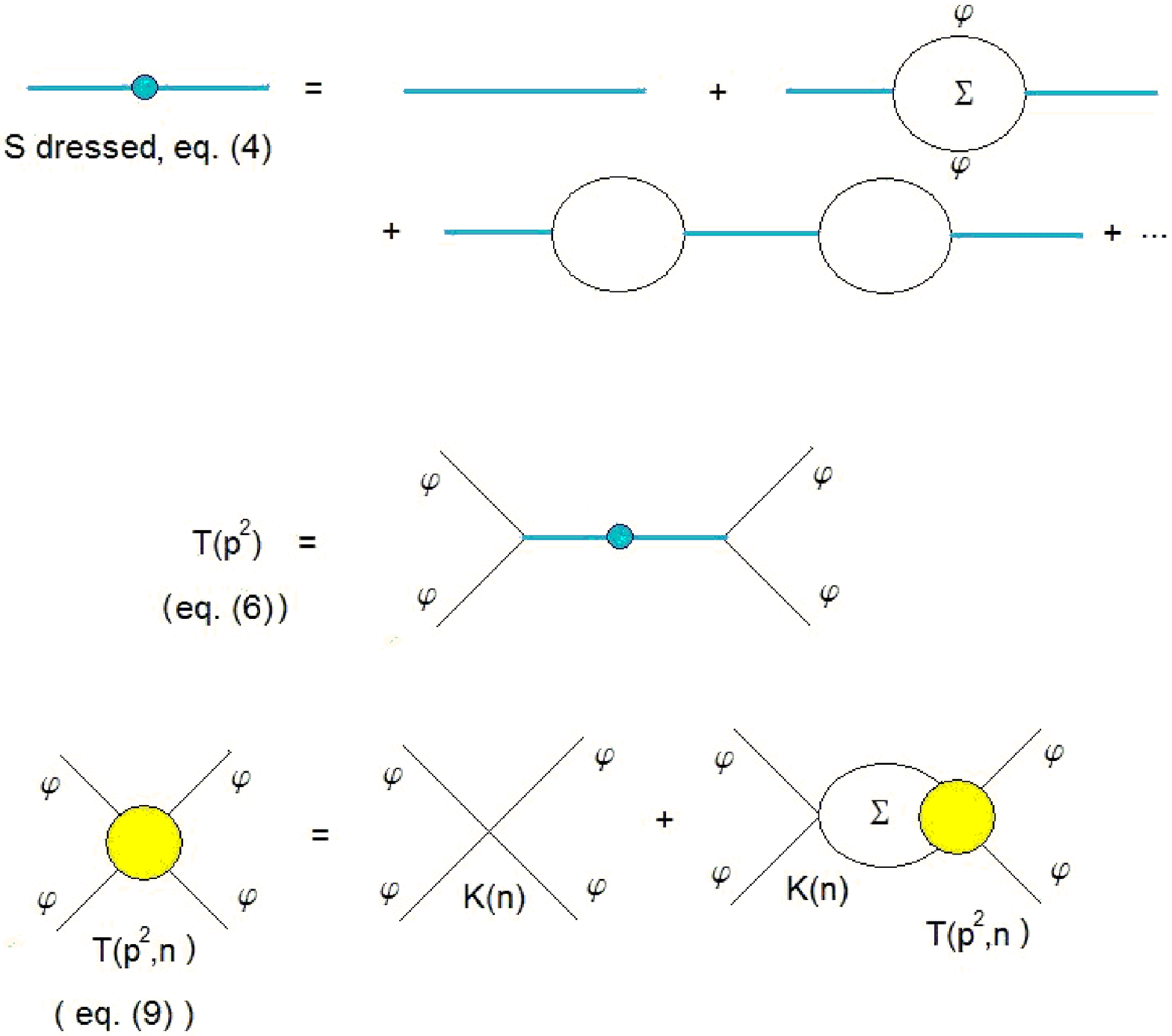}%
\caption{Eqs. (4), (6) and (9) are depicted: In eq. (4) the $S$ resonance is
dressed via loops of $\varphi$ mesons. In eq. (6) the $T$-matrix is
represented by an exchange of a dressed $S$ meson. Finally, eq. (9) represents
a BS equation applied to the quartic terms of the Lagrangian $\mathcal{L}%
_{\text{le}}(E_{\text{le}},N_{c})$ is represented, in which the $T$ matrix
appears both on the left and the right hand sides of the equation.}%
\end{center}
\end{figure}
%EndExpansion

The propagator of the field $S$ is modified via $\varphi$-meson loops and
takes the form (at the resummed 1-loop level, see Fig. 1):%
\begin{equation}
\Delta=i\left[  p^{2}-M_{0}^{2}+(\sqrt{2}g)^{2}\Sigma_{\Lambda}(p^{2})\right]
^{-1}%
\end{equation}
where $\Sigma_{\Lambda}(p^{2})$ is the 1-loop contribution, which is
regularized via a 3-d sharp cutoff $\Lambda$ \cite{fn3}. The dressed mass can
be defined via the zero of the real part of $\Delta^{-1},$ i.e.
\begin{equation}
M^{2}-M_{0}^{2}+(\sqrt{2}g)^{2}\operatorname{Re}\Sigma_{\Lambda}(M^{2})=0.
\end{equation}
In the large-$N_{c}$ limit $M\rightarrow M_{0}.$ This is true whatever
definition of the mass of the resonance is chosen. For finite $N_{c}$ one has
in general $M<M_{0}$ due to the loop corrections (see ref. \cite{lupo} for
details). The $T$ matrix for $\varphi\varphi$ scattering in the $s$-channel
upon 1-loop resummation is depicted in Fig. 1 and reads \cite{fn3primo}:%
\begin{equation}
T(p^{2})=i(\sqrt{2}g)^{2}\Delta=\frac{1}{-K^{-1}+\Sigma_{\Lambda}(p^{2}%
)},\text{ }K=\frac{(\sqrt{2}g)^{2}}{M_{0}^{2}-p^{2}}.\label{fullt}%
\end{equation}
Note, the present interest is focused on the 1-particle pole of the $S$
resonance and its corresponding enhancement in the $T$-matrix. Thus, for
simplicity we limit the study of two-body scattering to the exchange on one
(dressed) meson $S$. Other diagrams, as for instance the exchange of two (or
more) $S$ mesons, are not considered here (see, for instance, ref.
\cite{basdevant}). A more refined and complete approach should also include
the dressing of the $\varphi$ propagator and of the $S\varphi^{2}$ vertex. All
these complications, while important in a realistic treatment, can be
neglected at this illustrative level.

We now turn to development of a low-energy Lagrangian which involves only the
light meson field $\varphi.$ We assume that $g_{0}$ is not too large so that
for $N_{c}=3$ the mass $M$ lies above the threshold $2m.$ In this way it is
allowed to integrate out the field $S$ from $\mathcal{L}_{\text{toy}}$ and
obtain a low-energy (le) effective Lagrangian for the $\varphi\varphi$
interaction valid up to $E_{\text{le}}\lesssim2m<M_{0}$ \cite{leo}:%
\begin{equation}
\mathcal{L}_{\text{le}}(E_{\text{le}},N_{c})=-\frac{1}{2}\varphi(\square
+m^{2})\varphi+V,\text{ }V=\sum_{k=0}^{\infty}V^{(k)}, \label{le}%
\end{equation}%
\begin{equation}
\text{ }V^{(k)}=L^{(k)}\varphi^{2}\left(  -\square\right)  ^{k}\varphi
^{2},\text{ }L^{(k)}=\frac{g^{2}}{2M_{0}^{2+2k}}. \label{lec}%
\end{equation}
The Lagrangian $\mathcal{L}_{\text{le}}(E_{\text{le}},N_{c})$ contains only
quartic term of the kind $\varphi^{4},$ $\varphi^{2}\square\varphi^{2},$ ... .
$\mathcal{L}_{\text{le}}(E_{\text{le}},N_{c})$ is the analogue of chiral
perturbation theory or, more in general, of a low-energy Lagrangian in this
simplified system. The fact that we know explicitly the form of $\mathcal{L}%
_{\text{toy}}(E_{\text{max}},N_{c})$ allows us to calculate the `low-energy
constants' $L^{(k)}$ of eq. (\ref{lec}). If the form of $\mathcal{L}%
_{\text{toy}}(E_{\text{max}},N_{c})$ were unknown, then also the $L^{(k)}$
would be such. Note, each $L^{(k)}$ scales as $N_{c}^{-1}.$

\emph{BS-inspired unitarization, way 1}$\emph{:}$ As a first exercise let us
consider the low-energy Lagrangian $\mathcal{L}_{\text{le}}$ up to a certain
order $n$ by approximating the potential to $V(n)=\sum_{k=0}^{n}V^{(k)}$. By
performing a Bethe-Salpeter study with this approximate potential, see Fig. 1,
we obtain the following $T$ matrix:%
\begin{equation}
T(p^{2},n)=-K(n)+K(n)\Sigma_{\Lambda}(p^{2})T(p^{2},n) \label{bs}%
\end{equation}%
\begin{align}
T(p^{2},n)  &  =\frac{1}{-K(n)^{-1}+\Sigma_{\Lambda}(p^{2})},\text{
}\nonumber\\
K(n)  &  =\frac{(\sqrt{2}g)^{2}}{M_{0}^{2}}\sum_{k=0}^{n}\left(  \frac{p^{2}%
}{M_{0}^{2}}\right)  ^{k}. \label{tappr}%
\end{align}
where $K(n)$ is the bare tree-level amplitude corresponding to the sum of all
the quartic terms up to order $n.$

Clearly, $T(p^{2},n)$ is an approximate form of $T(p^{2})$ of eq.
(\ref{fullt}). The larger $n$, the better the approximation. Formally one has
$\lim_{n\rightarrow\infty}$ $T(p^{2},n)=T(p^{2})$. What we are doing is a
`dynamical reconstruction' of the state $S$ via a Bethe-Salpeter analysis
applied to the low-energy Lagrangian $\mathcal{L}_{\text{le}}$: We reobtain
the state $S$ which has been previously integrated out.

Let us keep $n$ fixed and perform a large-$N_{c}$ study of $T(p^{2},n).$ Do we
obtain the correct result, that is $M=M_{0}?$ The answer is no. In fact, in
the large-$N_{c}$ limit $K(n)$ scales as $1/N_{c}$ due to dependence encoded
in $g$, while $\Sigma_{\Lambda}(p^{2})$ scales as $N_{c}^{0}$ (we assume that
the cutoff does not scale with $N_{c}$ \cite{lamda}). In the large-$N_{c}$
limit we obtain $T(p^{2},n)\simeq-K(n).$ But $K(n)$ is a polynomial in $p^{2}$
and, for any finite $n,$ does not admit poles for finite $p^{2}$, but only for
$p^{2}\rightarrow\infty$. Thus, we find the incorrect result that in the
large-$N_{c}$ limit the mass of the dynamically reconstructed state is
infinity. This is shown in Fig. 2 for a particular numerical choice.%

%TCIMACRO{\FRAME{ftbpFU}{3.4022in}{2.5218in}{0pt}{\Qcb{Solid line: Absolute
%value of the full solution of the $T$ matrix $\left\vert T(x)\right\vert $
%with $x=\sqrt{p^{2}}$, eq. (\ref{fullt}). Dashed line: the approximate
%solution for $n=10,$ $\left\vert T(x,10)\right\vert $, eq. (\ref{tappr}). The
%values (GeV) $g_{0}=\Lambda=1.5$, $m=0.3$ and $M_{0}=1$ are used. The dressed
%mass reads $M=0.96$ . The agreement is very good up to 1 GeV for $N_{c}=3$
%(panel (a)). When increasing $N_{c}$ the full solution is centered on $M_{0}$
%and becomes narrower, as it should. On the contrary, the peak of the
%approximate solution increases and the width is only slightly affected by it.
%The approximate solution does not have the correct large-N$_{c}$ expected
%behavior.}}{\Qlb{fig1}}{fig1.eps}{\special{ language "Scientific Word";
%type "GRAPHIC";  display "USEDEF";  valid_file "F";  width 3.4022in;
%height 2.5218in;  depth 0pt;  original-width 5.2434in;
%original-height 8.2538in;  cropleft "0";  croptop "1";  cropright "1";
%cropbottom "0";  filename 'fig1.eps';file-properties "XNPEU";}} }%
%BeginExpansion
\begin{figure}
[ptb]
\begin{center}
\includegraphics[
height=2.5218in,
width=3.4022in
]%
{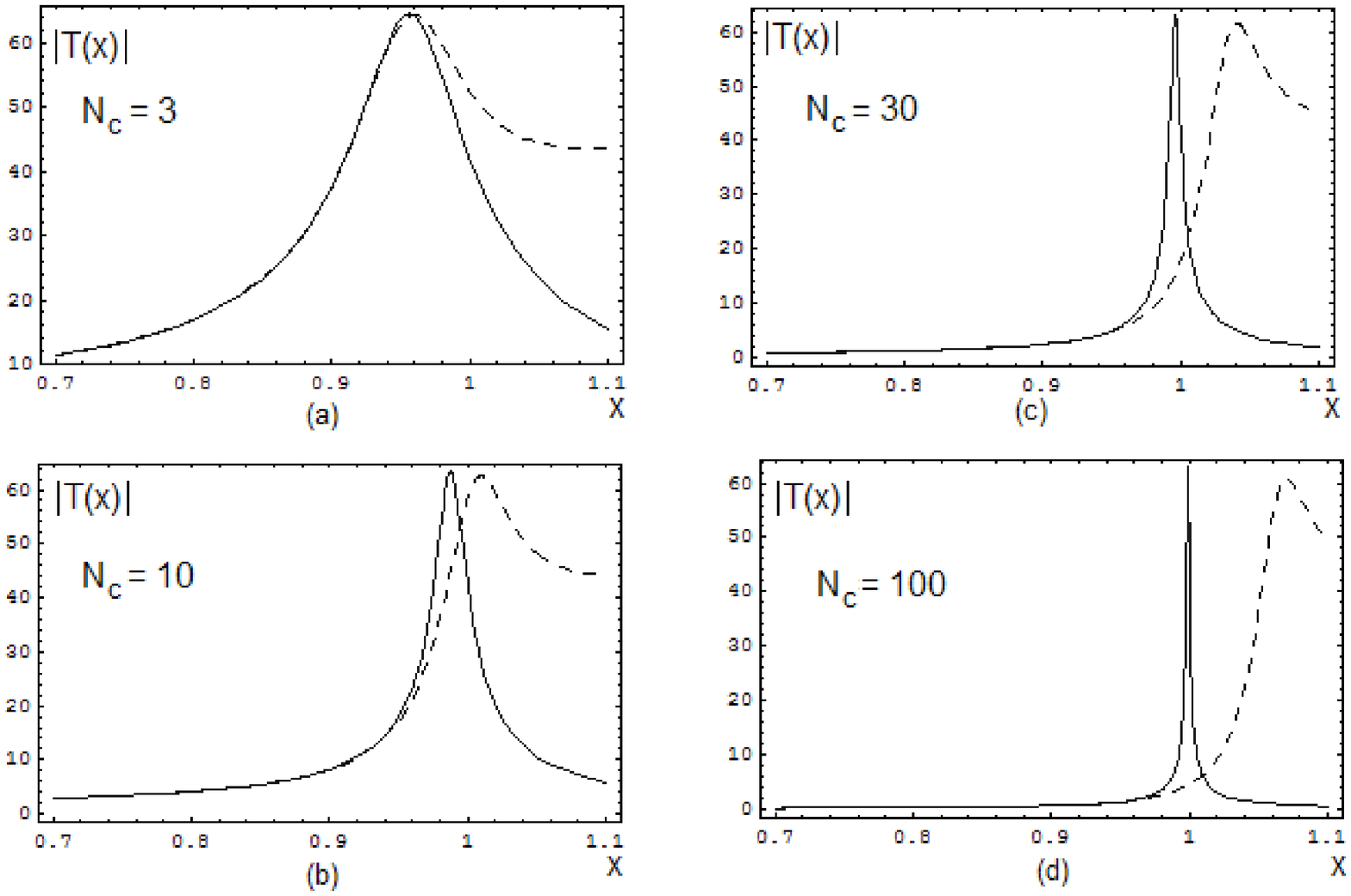}%
\caption{Solid line: Absolute value of the full solution of the $T$ matrix
$\left\vert T(x)\right\vert $ with $x=\sqrt{p^{2}}$, eq. (\ref{fullt}). Dashed
line: the approximate solution for $n=10,$ $\left\vert T(x,10)\right\vert $,
eq. (\ref{tappr}). The values (GeV) $g_{0}=\Lambda=1.5$, $m=0.3$ and $M_{0}=1$
are used. The dressed mass reads $M=0.96$ . The agreement is very good up to 1
GeV for $N_{c}=3$ (panel (a)). When increasing $N_{c}$ the full solution is
centered on $M_{0}$ and becomes narrower, as it should. On the contrary, the
peak of the approximate solution increases and the width is only slightly
affected by it. The approximate solution does not have the correct
large-N$_{c}$ expected behavior.}%
\label{fig1}%
\end{center}
\end{figure}
%EndExpansion

Although our analysis has been applied to a simple toy model, the form of eq.
(\ref{bs}) is general. One has a polynomial form for $K(n)$ as function of
$p^{2}$ and a mesonic loop $\Sigma_{\Lambda}(p^{2})$ which is independent on
$N_{c}.$ Complications due to different quantum numbers do not alter the
conclusion. We also note that these results are in agreement with the
discussion of ref. \cite{zheng}, where the scalar $\sigma$ meson is first
integrated out and then `reconstructed' in the framework of the linear sigma model.

\emph{IAM inspired unitarization }If we would, instead, apply the IAM
unitarization scheme to the $n=1$ approximate form we would obtain the correct
result in the large-$N_{c}$ expansion. In fact, in this case one schematically
has (neglecting $t$ and $u$ channels):
\begin{equation}
T_{\text{IAM}}\simeq T_{2}\left(  T_{2}-T_{4}-iT_{2}\sigma T_{2}\right)
^{-1}T_{2}%
\end{equation}
where $\sigma=\sqrt{\frac{p^{2}}{4}-m^{2}}$ in our notation. Being
$T_{2}=\frac{(\sqrt{2}g)^{2}}{M_{0}^{2}}$ and $T_{4}=-\frac{(\sqrt{2}g)^{2}%
}{M_{0}^{4}}p^{2}$ one finds
\begin{equation}
T_{\text{IAM}}\simeq(\sqrt{2}g)^{2}\left(  M_{0}^{2}-p^{2}-i(\sqrt{2}%
g)^{2}\sigma\right)  ^{-1}%
\end{equation}
which represents a valid approximation of the full $T$ matrix if $g$ is not
too large $(M\simeq M_{0}).$ It is straightforward to see that the IAM
approximation delivers the correct large-$N_{c}$ result, namely $M\rightarrow
M_{0}$ and a width decreasing as $1/N_{c}.$ Clearly one could repeat this
study for increasing $n$ finding a better and better approximation of $T$.

\emph{BS-inspired unitarization, way 2}$\emph{:}$\emph{ }Contrary to the
BS-inspired unitarization described above (way 1), it is possible to follow a
different BS-inspired approach which is in agreement with the large-$N_{c}$
limit. For simplicity we discuss it in the explicit case $n=1$ \cite{fnthank}.
One has:
\begin{equation}
K(1)=\frac{(\sqrt{2}g)^{2}}{M_{0}^{2}}\left(  1+\frac{p^{2}}{M_{0}^{2}%
}\right)  .
\end{equation}
Now, instead of plugging $K(1)^{-1}$ directly into eq. (\ref{tappr}), we first
invert it obtaining the approximate form $K(1)_{\text{way2}}^{-1}$ valid up to
order $O(p^{4}/M_{0}^{4})$:
\begin{equation}
K(1)_{\text{way2}}^{-1}=\frac{M_{0}^{2}}{(\sqrt{2}g)^{2}}\left(  1-\frac
{p^{2}}{M_{0}^{2}}+...\right)
\end{equation}
The next step is to write the $T$ matrix in terms of $K(1)_{\text{way2}}^{-1}%
$:
\begin{align}
T(p^{2},1)_{\text{way2}}  &  =\frac{1}{-K(1)_{\text{way2}}^{-1}+\Sigma
_{\Lambda}(p^{2})}\\
&  \simeq\frac{(\sqrt{2}g)^{2}}{p^{2}-M_{0}^{2}+(\sqrt{2}g)^{2}\Sigma
_{\Lambda}(p^{2})}.
\end{align}
Thus, this new approximate form derived from the BS equation is now in
agreement with the large-$N_{c}$ limit and is equivalent to the
IAM-inspired\ unitarization approach described above. This shows an important
fact in this discussion: it is not the BS method which fails in BS-way 1, but
rather the adopted perturbative expansion. We could as well develop a second
IAM-inspired unitarization which fails to reproduce the correct large $N_{c}$
results and that is equivalent to BS-way1. From this perspective we can
rearrange the unitarizations as `large $N_{c}$ correct' (BS-way2 and IAM) and
`large $N_{c}$ violating' (BS-way1 and IAM-way2). The reason why we associate
to the the different unitarizations the names BS or IAM is simply due to the
way the equations are settled down in the different cases. It offers a simple
mnemonic to their development.

By studying the large-$N_{c}$ limit one can see closer the relations between
the two described BS-unitarizations: in the case $n=1$ and in the
large-$N_{c}$ limit, the $T$ matrix in the first BS form reads $T_{\text{way1}%
}\simeq-K(1)=-\frac{(\sqrt{2}g)^{2}}{M_{0}^{2}}\left(  1+\frac{p^{2}}%
{M_{0}^{2}}\right)  ,$ which obviously has no pole. In the second
BS-unitarization one has in the large-$N_{c}$ limit $T_{\text{way2}}%
\simeq-K(1)_{\text{way2}}\simeq-\frac{(\sqrt{2}g)^{2}}{M_{0}^{2}}\left(
1-\frac{p^{2}}{M_{0}^{2}}\right)  ^{-1}$ and the correct pole $p^{2}=M_{0}%
^{2}$ is recovered.

It is however important to notice that -just as in the IAM\ case- at least two
terms in the expansion of the amplitude $K$ are necessary to perform this
second BS unitarization. This is the reason why it cannot be applied in the
case studied in the next subsection (Sec. IV.B), where only the lowest term of
the amplitude is kept.

\subsection{BS equation with the lowest term only}

In most studies employing the BS analysis only the lowest term of the
effective low-energy Lagrangian is kept. Within the present toy model it is
not possible to reconstruct a resonance with mass $M>2m$ with only the lowest
term ($n=0$) \cite{fn4}. However, a simple modification of the model which
allows for such a study is possible:
\begin{equation}
\mathcal{L}_{\text{toy}}^{\text{new}}(E_{\text{max}},N_{c})=\mathcal{L}%
_{\text{toy}}(E_{\text{max}},N_{c})+\frac{g^{2}}{2M_{0}^{2}}\varphi^{4}.
\label{lnew}%
\end{equation}
In this way an extra-repulsion (whose quartic form is assumed to be valid up
to $E_{\text{max}}$) has been introduced. The $T$ matrix takes the form:
\begin{equation}
T(p^{2})=\frac{1}{-K^{-1}+\Sigma_{\Lambda}(p^{2})},\text{ }K=\frac{(\sqrt
{2}g)^{2}}{M_{0}^{2}-p^{2}}-\frac{(\sqrt{2}g)^{2}}{M_{0}^{2}} \label{fullt2}%
\end{equation}
When deriving the low-energy Lagrangian everything goes as before, but the
$k=0$ term is now absent:%
\[
V(n)=\sum_{k=1}^{n}V^{(k)},\text{ }V^{(k)}=L^{(k)}\varphi^{2}\left(
-\square\right)  ^{k}\varphi^{2}%
\]
Note, in this case the $\varphi\varphi$ scattering vanish at low momenta and
in the chiral limit $m\rightarrow0$ (just as the $\pi\pi$ scattering in reality).

A study of the case $n=1$ (corresponding to the first term only in the
expansion) is now possible. We consider the following situation: Let the
original Lagrangian $\mathcal{L}_{\text{toy}}^{\text{new}}$ of eq.
(\ref{lnew}) be unknown. The low-energy potential at the lowest order reads
$V\simeq$ $V^{(1)}=L^{(1)}\varphi^{2}\left(  -\square\right)  \varphi^{2}$,
but the low-energy coefficient $L^{(1)}$ is also unknown. Moreover, from low
energy informations only one does not know the value of the cutoff $\Lambda$
to be employed in mesonic loops: a new cutoff $\widetilde{\Lambda}$, not
necessarily equal to the original $\Lambda,$ is also introduced as a free
parameter. From the perspective of low-energy phenomenology, one writes down
the following approximate form for the $T$ matrix, which depend on two `free
parameters' $L^{(1)}$ and $\widetilde{\Lambda}$:
\begin{align}
\widetilde{T}(p^{2})  &  =T(p^{2},1)=\frac{1}{-\widetilde{K}^{-1}%
+\Sigma_{\widetilde{\Lambda}}(p^{2})},\text{ }\nonumber\\
\widetilde{K}  &  =K(1)=4L^{(1)}p^{2}. \label{tappr2}%
\end{align}
The question is if it is possible to vary $L^{(1)}$ and $\widetilde{\Lambda}$
in such a way that the approximate $T$-matrix $\widetilde{T}(p^{2})$
reproduces the `full' result $T(p^{2})$ of eq. (\ref{fullt}) between, say,
$2m=0.6$ GeV and $1.3$ GeV for $N_{c}=3$.%

%TCIMACRO{\FRAME{ftbpFU}{3.5068in}{2.1837in}{0pt}{\Qcb{ Full $\left\vert
%T(x)\right\vert $ (solid line, eq. (\ref{fullt2}) with $\Lambda=1.5$ GeV) and
%approximate $\left\vert \widetilde{T}(x)\right\vert $ (dashed line, eq.
%(\ref{tappr2})) in the cases $g_{0}=1.5$ (left column) and $g_{0}=5$ (right
%column) GeV for different values of $N_{c}.$ The values of $L^{(1)}$ and
%$\widetilde{\Lambda}$ , which determine the approximate dashed curve, are
%determined by fitting the approximate form to the full one in the $N_{c}=3$
%cases. One has $\widetilde{\Lambda}\simeq15000\Lambda$ in the left column, and
%$\widetilde{\Lambda}=\Lambda$ in the right column. As soon as $N_{c}$ is
%increased the approximate solution quickly fades out, while the real solution
%approaches $M_{0}$ where it becomes more and more peaked.}}{\Qlb{fig2}%
%}{fig2.eps}{\special{ language "Scientific Word";  type "GRAPHIC";
%width 3.5068in;  height 2.1837in;  depth 0pt;  original-width 9.5562in;
%original-height 5.9326in;  cropleft "0";  croptop "1";  cropright "1";
%cropbottom "0";  filename 'fig2.eps';file-properties "XNPEU";}} }%
%BeginExpansion
\begin{figure}
[ptb]
\begin{center}
\includegraphics[
height=2.1837in,
width=3.5068in
]%
{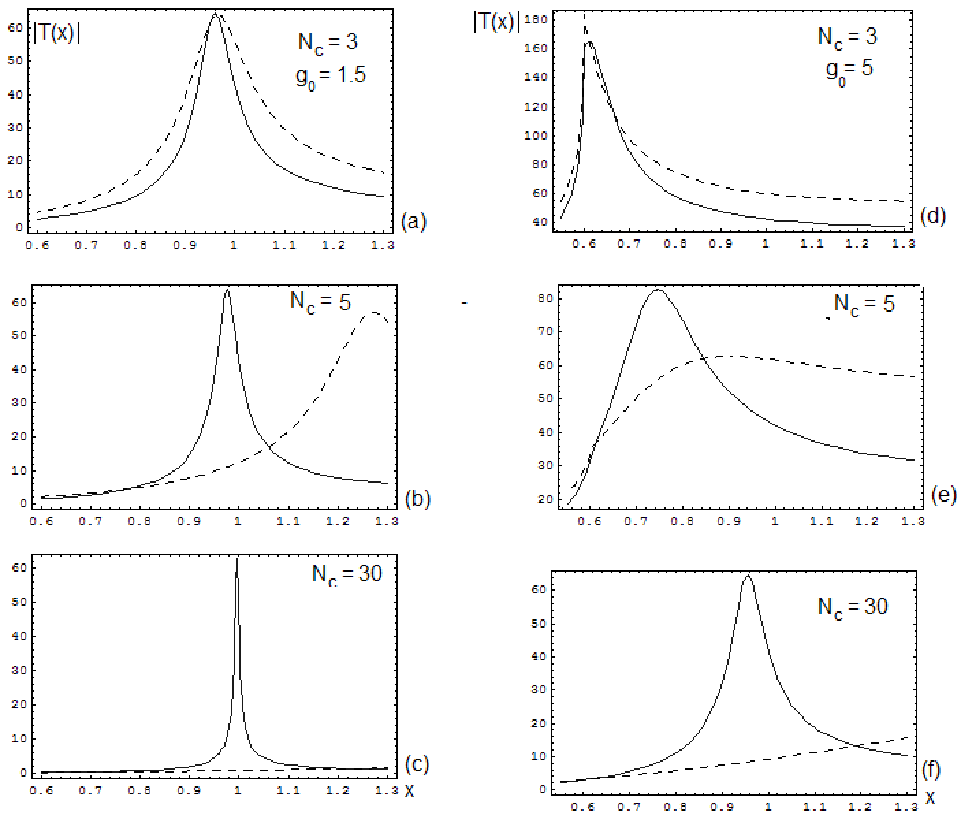}%
\caption{ Full $\left\vert T(x)\right\vert $ (solid line, eq. (\ref{fullt2})
with $\Lambda=1.5$ GeV) and approximate $\left\vert \widetilde{T}%
(x)\right\vert $ (dashed line, eq. (\ref{tappr2})) in the cases $g_{0}=1.5$
(left column) and $g_{0}=5$ (right column) GeV for different values of
$N_{c}.$ The values of $L^{(1)}$ and $\widetilde{\Lambda}$ , which determine
the approximate dashed curve, are determined by fitting the approximate form
to the full one in the $N_{c}=3$ cases. One has $\widetilde{\Lambda}%
\simeq15000\Lambda$ in the left column, and $\widetilde{\Lambda}=\Lambda$ in
the right column. As soon as $N_{c}$ is increased the approximate solution
quickly fades out, while the real solution approaches $M_{0}$ where it becomes
more and more peaked.}%
\label{fig2}%
\end{center}
\end{figure}
%EndExpansion
The answer is that this is generally possible, but the results for $L^{(1)}$
and $\widetilde{\Lambda}$ vary drastically with the coupling constant $g_{0}$
in the original Lagrangian. In particular, if $g_{0}$ is small a good fit
implies a very large and unnatural value of $\widetilde{\Lambda}$ (Fig. 3.a).
For instance, for $g_{0}=g(N_{c}=3)=1.5$ GeV the mass $M=0.95$ GeV is only
slightly shifted from the bare mass $M_{0}=1$ GeV. In this case the
approximate form $\left\vert \widetilde{T}(p^{2})\right\vert $ reproduces
$\left\vert T(p^{2})\right\vert $ only if $\widetilde{\Lambda}$ $\sim
10^{4}\Lambda$ (astronomically high and seemingly unnatural from the
perspective of the low-energy theory!).

The situation changes completely if $g_{0}$ is large: it is possible to find a
satisfactory description in which $\widetilde{\Lambda}\sim\Lambda$. For
instance, for $g_{0}=5$ GeV one has $M=0.65$ GeV and the approximate
$\left\vert \widetilde{T}(p^{2})\right\vert $ reproduces well $\left\vert
T(p^{2})\right\vert $ for $\widetilde{\Lambda}=\Lambda$ (Fig. 3.d)

In the first case the failure of the dynamical reconstruction with a
meaningful value of the cutoff $\widetilde{\Lambda}$ is due to the
quantitative inappropriate behavior of the Bethe-Salpeter approach when only
the first term is kept. In the second case a rather satisfactory description
is possible for a meaningful value of the cutoff. On the light of the results
of the low-energy Lagrangian only, one could also propose the interpretation
that the obtained state $S$ is dynamically generated, and shall be regarded as
a $\varphi\varphi$ molecular state. This is, however,\emph{ not} the correct
interpretation in the present example. We know, in fact, that this state
corresponds -by construction- to the original, preexisting, quarkonium-like
state $S$.

In both cases, as soon as we increase the number of colors, the approximate
$T$-matrix $\left\vert \widetilde{T}(p^{2})\right\vert $ and the full
$\left\vert T(p^{2})\right\vert $ show a completely different behavior (Fig.
3, panels (b,c) and (e,f)): While the peak of $\left\vert T(p^{2})\right\vert
$ approaches $M_{0}=1$ GeV and becomes narrower according to the correct
large-$N_{c}$ limit of the $S$ meson, the dynamically reconstructed state
fades out, because of the\emph{ incorrect} behavior of BS unitarization with
large-$N_{c}.$ This is clearly visible from the interaction term
$V^{(1)}=L^{(1)}\varphi^{2}\left(  -\square\right)  \varphi^{2},$ because
$L^{(1)}$scales as $N_{c}^{-1}$. However, although the interaction term
disappears with large-$N_{c},$ the state $S$ is still the original
quark-antiquark state. This example shows that the reconstruction of the state
$S$ is not possible in the large-$N_{c}$ limit, but does not mean that $S$ is
a dynamically generate state of molecular type. Note, this is just a sub-case
of the previous general discussion on large-$N_{c}$ dependence: the fact that
only one term is kept generates a much faster `fading out' of the
reconstructed state, compare Fig. 2 and Fig. 3.

In the previous subsection it was shown that -while a straight application of
the BS equation is at odds with the large-$N_{c}$ limit- a second
BS-unitarization allows for a correct description of the large-$N_{c}$ limit.
The second BS-unitarization is however not applicable in the present case. In
fact, \emph{at least} two terms in the expansion of $K(n)$ are needed to
follow it. If only the lowest term is kept, as done here with the term $n=1$
in Eq. (\ref{tappr2}), this is no longer feasible. This is similar to the fact
that also the IAM method needs at least two terms in the expansion of the
amplitude $K$ in order to be applicable \cite{fnnew}.

\subsection{Analogy with the real world}

The original toy-Lagrangian $\mathcal{L}_{\text{toy}}(E_{\text{max}},N_{c})$
of eq. (\ref{ltoy}) is assumed to be valid up to an energy $E_{\text{max}%
}>>M_{0}.$ The corresponding low-energy Lagrangian $\mathcal{L}_{\text{le}%
}(E_{\text{le}},N_{c})$ of eq. (\ref{le}) -obtained by integrating out the $S$
field- is valid up to an energy $E_{\text{le}}<<M_{0}.$ When unitarizing
$\mathcal{L}_{\text{le}}(E_{\text{le}},N_{c})$, one can enlarge the validity
of the low-energy theory up to $M_{0}$ and then infer the existence of the
resonance $S$ with mass $M<M_{0}.$ However, if no other input is known, the
nature of the state $S$ cannot be further studied, see the general discussion
of the point (d) in Sec. II.

This situation is similar to the example (c) in\ Sec. II: the Fermi Lagrangian
$\mathcal{L}_{\text{F}}$ alone does not allow to deduce the nature of the $W$
meson, even if the existence of the latter is inferred by unitarization
arguments applied to $\mathcal{L}_{\text{F}}$. It is also similar to the cases
studied in Sec. III.C: when a resonance is obtained by unitarizing a
low-energy mesonic Lagrangian, (at first) no statement about its nature can be done.

Further information is needed: in the case of the $W$ meson the full
electroweak Lagrangian $\mathcal{L}_{\text{EW}}$ is known and leads to the
straightforward conclusion that the $W$ meson is not a dynamically generated
state, but a fundamental field of the standard model. In the framework of the
toy model, this corresponds to the knowledge of the `full Lagrangian'
$\mathcal{L}_{\text{toy}}(E_{\text{max}},N_{c})$ of eq. (\ref{ltoy}). The
`quarkonium' nature of $S$ can then be easily deduced.

In the case of low-energy mesonic theories discussed in Sec. III.C, the full
hadronic Lagrangian is not known. The only additional knowledge is the
large-$N_{c}$ scaling of the low-energy constants of the low-energy
Lagrangian(s). In the framework of the toy model, this corresponds to the
knowledge of low-energy Lagrangian $\mathcal{L}_{\text{le}}(E_{\text{le}%
},N_{c})$ of eq. (\ref{le}) (up to a certain $n$) together with the scaling of
the quantities $L^{(k)}$ in eq. (\ref{lec}). The latter additional knowledge
can lead to the correct conclusions about the nature of the $S$ meson,
although -as discussed in Sec. IV.A- care is needed when the BS method is chosen.

Moreover, as further studied in Section. IV.B, when only \emph{the lowest
term} of the low-energy Lagrangian is kept, it is not possible to reproduce
the correct large-$N_{c}$ behavior of the resonance $S.$ Although the
`dynamical reconstruction' of the state $S$ is possible, the state $S$ `looks
like' a molecular state which fades out in the large-$N_{c}$ limit. This
however is not true: In fact, we know from the very beginning that the state
$S$ corresponds `by construction' to a quark-antiquark state.

Although this discussion is based on toy models and the real world is much
more complicate than this, the same qualitative picture can hold in low-energy
QCD. In fact, the use of the BS equation in the Literature is often limited to
the lowest term of a low-energy Lagrangians for $\pi\pi,$ $\rho\rho,$ ...
interactions. In our view, such low-energy Lagrangians emerge upon integrating
out all the heavier fields in $\mathcal{L}_{eff}^{\text{had}}(E_{\text{max}%
},N_{c}=3),$ in which tensor, axial-vector and scalar quark-antiquark fields
must exist below 2 GeV. Then, the use of the BS equation, similarly to the
dynamical reconstruction of $S$ in this simple example, leads to the dynamical
reconstruction of the axial-vector, tensor and scalar mesons above 1 GeV: they
are preexisting, quark-antiquark states, which are reobtained from low-energy
Lagrangians via unitarization methods. Future unitarization studies, involving
the leading and the next-to-leading-terms in the effective Lagrangians, may
shed light on this point.

\section{Conclusions and outlook}

In this work we studied the issue of dynamical generation both in a general
context and in low-energy QCD. A dynamically generated resonance has been
defined as a state which does not correspond to any of the fields of the
original Lagrangian describing the system up to a certain maximal energy
$E_{\text{max}}$, provided that its mass lies below this maximal energy. This
discussion also offered us the possibility to distinguish in principle
tetraquark from mesonic molecular states in low-energy QCD: while the former
are fundamental and shall be included as bare fields in the (yet-unknown)
hadronic Lagrangian $\mathcal{L}_{eff}^{\text{had}}(E_{\text{max}},N_{c}=3)$,
this is not the case for the latter. Note, $\mathcal{L}_{eff}^{\text{had}%
}(E_{\text{max}},N_{c}=3)$ represents the complete hadron theory valid up to
$E_{\text{max}}\simeq2$ GeV.

In the application to the hadronic world we also discussed dynamical
reconstruction of resonances: these are resonances which are obtained via
unitarization methods from low-energy effective Lagrangians, but still
represent fundamental fields (such as quark-antiquark states) in
$\mathcal{L}_{eff}^{\text{had}}(E_{\text{max}},N_{c}=3)$. Note, the low-energy
effective Lagrangians can be seen as the result of integrating out heavier
(quarkonia, glueballs, ...) fields representing intrinsic, fundamental states
in $\mathcal{L}_{eff}^{\text{had}}(E_{\text{max}},N_{c}=3)$. In the scenario
of dynamical reconstruction, one reconstructs these heavier resonances by
unitarizing the appropriate low-energy Lagrangian.

Within a simple toy model these issues have been examined. This model consists
of two fields, $\varphi$ and $S,$ with the latter being heavier and with a
nonzero decay width into $\varphi\varphi.$ We introduced a large-$N_{c}$
dependence which mimics that of quarkonium states in QCD. The field $S$ has
been first integrated out and the emerging low-energy interaction Lagrangian
involving only the field $\varphi$ has been derived. Out of it the state $S$
has been `dynamically reconstructed'.

In order to do it we have used a unitarization inspired by the Bethe-Salpeter
equation and we have shown that the original, quarkonium-like large-$N_{c}$
behavior of $S$ cannot be reproduced if only the lowest term in the effective
Lagrangian is kept. (Note, when more terms are kept this problem can be easily
solved and the large $N_{c}$ result is correct also within the BS approach.
The problem is not the latter but the adopted perturbative expansion, see Sec
IV.A). We then proposed that a similar, although more complicated, dynamical
reconstruction mechanism takes place for tensor, axial-vector and scalar
mesons above 1 GeV: these resonances, studied in recent works, can be
interpreted as fundamental quark-antiquark states, which are reobtained when
unitarizing low-energy effective Lagrangians. In this scenario there is no
conflict between the `old' quark model assignments and recent developments,
because they would both represent a dual description of the same, preexisting
quark-antiquark resonances. This interpretation, although not yet conclusive,
represents a possibility which deserves further study.

Dynamical reconstruction can also hold for light scalar mesons below 1 GeV, if
the latter form a quarkonium (quite improbable) or a tetraquark nonet. The
situation in this case is, as discussed in the text, still unclear. In this
work we limited the study to the light mesonic sector, but the present
discussion about dynamical generation/reconstruction can also hold, with due
changes, in the baryon and in the heavy quark sectors.

\bigskip

%%%%%%%%%%%%%%%%%%%%%%%%%%%%%%%%%%%%%%%%%%%%%%%%%%%%%%%%%%%%%%%%%%%%%%%%
\textbf{Acknowledgements} The author thanks D. H. Rischke, S. Leupold and T.
Brauner for useful discussions. Financial support from BMBF is acknowledged.
%%%%%%%%%%%%%%%%%%%%%%%%%%%%%%%%%%%%%%%%%%%%%%%%%%%%%%%%%%%%%%%%%%%%%%


\begin{thebibliography}{99}
\bibitem {fnlast}The definition of Sec. II does not allow to conclude that the
resonance $R$ obtained in the framework of unitarized $\mathcal{L}$ is
dynamically generated because $M_{R}>$ $E_{\text{max}}.$ This is in agreement
with the fact that at the level of the unitarized version of $\mathcal{L}$ one
cannot deduce if the field $R$ appears explicitly or not in the more complete
Lagrangian $\mathcal{L}^{\prime}$ (i.e. one cannot state if $R$ is dynamically
generated or not). Note, even in the case $E_{U}\sim E_{\text{max}}^{\prime},$
the unitarized version of $\mathcal{L}$ is not equivalent to $\mathcal{L}%
^{\prime}$ (therefore one should not use the definition of Sec. II with the
energy scale $E_{U}$ instead of $E_{\text{max}}$). Only some of the properties
of $\mathcal{L}^{\prime}$ can be inferred by the extrapolation to higher
energy of the information contained in the Lagrangian $\mathcal{L}$.
\bibitem {fn5}The theory is non-renormalizable because of the logarithmic term
in $V_{G}$. Moreover, as soon as (axial-)vector fields are included, further
non-renormalizable terms emerge.
\bibitem {fn3}The details of the regularization are not important, as long as
momenta larger than $\Lambda$ are suppressed in the meson loop. Note that,
although the toy model is formally renormalizable, we still work with a finite
cutoff which directly affects the results. This is namely also the situation
in the true world: a cutoff of the order of 1 GeV is an essential ingredient
in many effective models of low-energy QCD.
\bibitem {lamda}The assumption that $\Lambda$ does not depend on $N_{c}$ can
be justified by assuming that $\Lambda\sim\Lambda_{QCD},$ the latter being
$N_{c}$ independent. An alternative behavior of $\Lambda$ as function of
$N_{c}$ have been discussed in ref. \cite{avnc}, where the case $\Lambda
\sim4\pi f_{\pi}$ leads to an increase of $\Lambda$ with $N_{c}^{1/2}.$ Here
we note that in the context of the toy model the mass of the reconstructed
state still diverges when admitting a power-like dependence $\Lambda\sim
N_{c}^{\alpha}$ with $\alpha\geq0,$ although the quantitative way how the
limit is reached is strongly affected by the value of $\alpha.$
\bibitem {fn4}This is due to the fact that the lowest term has no $p^{2}$
dependence and to the particular form of $\operatorname{Re}\Sigma(p^{2})$, see
Ref. \cite{lupo}. The reconstructed mass is actually infinite. The first term
only, although inadequate in this case, could still work if the loops were so
strong that $M<2m$ (even if $M_{0}>2m$), but we do not discuss this case
here.
\bibitem {fnnew}If, instead of $n=1,$ the case $n=2$ is considered, then again
the second BS unitarization is possible. In fact, being $K(2)=\frac{(\sqrt
{2}g)^{2}p^{2}}{M_{0}^{4}}\left(  1+\frac{p^{2}}{M_{0}^{2}}\right)  $ one has
$K(2)_{\text{way2}}^{-1}=\frac{M_{0}^{4}}{(\sqrt{2}g)^{2}p^{2}}\left(
1-\frac{p^{2}}{M_{0}^{2}}+...\right)  $ and $T_{\text{way2}}\simeq\frac
{(\sqrt{2}g)^{2}(p^{2}/M_{0}^{2})}{p^{2}-M_{0}^{2}-(\sqrt{2}g)^{2}(p^{2}%
/M_{0}^{2})\Sigma_{\Lambda}}$, which is in agreement with the large-$N_{c}$
expectation.
\end{thebibliography}
\end{document}